\documentclass[prl,twocolumn,floats,aps,epsfig,nofootinbib,amssymb]{revtex4}
\usepackage[dvips]{graphicx}
\usepackage{amsmath}
\usepackage{bm}
\usepackage{times}
\usepackage{epsfig}
\usepackage{graphicx}
\usepackage{color}
\usepackage{cancel}
\usepackage{amssymb}
\usepackage{textcomp}
\begin{document}
\title{\Large Type III Seesaw and Left-Right Symmetry}
\bigskip
\author{Pavel Fileviez P{\'e}rez}
\email{fileviez@physics.wisc.edu}
\address{\\
University of Wisconsin-Madison, Department of Physics \\
1150 University Avenue, Madison, WI 53706, USA}
\date{\today}
\begin{abstract}
The implementation of the Type III seesaw mechanism for
neutrino masses in the context of left-right theories
where parity is broken spontaneously is investigated.
We propose a simple left-right symmetric theory where
the neutrinos masses are generated through a double
seesaw mechanism which is a combination of Type I
and Type III seesaw. In this context we find a possible
candidate for the cold dark matter in the Universe and
discuss the Baryogenesis via Leptogenesis mechanisms. 
The spectrum of the theory, the phenomenological constraints
and the possibility to test the theory at
the Large Hadron Collider are investigated.
\end{abstract}
\maketitle
\section{I. Introduction}
In the LHC Era we hope to test the theory beyond the Standard Model
(SM) which describes physics at the TeV scale. The existence of
massive neutrinos is a strong motivation for physics beyond
the SM. There are only three simple mechanisms to generate 
Majorana neutrino masses at tree level. In the case of Type I
seesaw mechanism~\cite{TypeI-1, TypeI-2}, one can
add at least two fermionic singlets $N_i$ (right-handed neutrinos)
and the neutrino masses read as $m_\nu^I \simeq h^2_\nu v^2/ M_N$,
where $h_\nu$ is the Dirac Yukawa coupling, $v=246$ GeV is the SM 
Higgs vacuum expectation value (vev) and $M_N$ is the 
right-handed neutrino mass. If $h_\nu \simeq 1$ and $M_{N} 
\approx 10^{14-15}$ GeV, one obtains the natural value 
for the neutrino masses $m_\nu \approx 1$ eV.

In the so-called Type II seesaw mechanism~\cite{TypeII} for 
neutrino masses the Higgs sector of the SM is extended by adding 
an $SU(2)_L$ Higgs triplet $\Delta$. In this scenario the neutrino
masses are given by $m_\nu^{II} \simeq Y_\nu  v_{\Delta}$, where
$v_{\Delta}$ is the vev of the neutral component of the triplet 
and $Y_\nu$ is the Yukawa coupling. $v_{\Delta} \simeq \mu v^2/M_{\Delta}^2$, 
where $M_{\Delta}$ is the mass of the triplet and $\mu$ defines 
the mixing between the SM Higgs and the triplet. A natural setting 
would be $Y_\nu \approx 1$ and $\mu \sim M_{\Delta} \approx 10^{14-15}$ GeV.

Recently, several groups have investigated the implementation of the
Type III seesaw mechanism~\cite{TypeIII, Ma, BS, P1} in the context of
grand unified theories. In this case adding at least two extra matter
fields in the adjoint representation of $SU(2)_L$ and with zero hypercharge,
one can generate neutrino masses, $m_\nu^{III} \simeq \Gamma^2_\nu v^2/M_\rho$. 
Here $M_\rho$ stands for the mass of the fermionic triplets and $\Gamma_\nu$ is 
the Dirac Yukawa coupling. The implementation of this mechanism~\cite{Ma, BS, P1} 
has been studied in the context of $SU(5)$ grand unified theories, where 
once we realize Type III seesaw, one gets Type I as a bonus since 
both fields responsible for seesaw live in the adjoint 
representation of $SU(5)$.

Parity is considered as a fundamental symmetry and is explicitly broken
in the SM by the asymmetry between the left and right
handed multiplets. Therefore, as is well known the SM does not explain
the $V-A$ character of the $\beta$ and $\mu$ decays. One could say
that the existence of massive neutrinos and the unknown origin of
parity violation in the SM are probably one of the main physical
motivations for physics beyond the SM. In the context of the
so-called left-right symmetric theories~\cite{Pati-Salam,Pati-Mohapatra}
one has the appealing possibility to understand the origin of parity violation and
its strong connection to the generation of neutrino masses~\cite{TypeI-2}.
In these theories the observed V-A structure of weak interactions is only
a low-energy phenomenon which should disappear when one reaches the
TeV scale or higher. Left-right symmetric theories where the neutrino masses
are generated through the Type I and Type II seesaw mechanisms have been
investigated in great detail~\cite{Deshpande:1990ip, Barenboim:2001vu, Zhang:2007da, GoranLR, Senjanovic-Mohapatra, Langacker}.

In this Letter we study for the first time the 
implementation of the Type III seesaw mechanism in 
the context of left-right theories. We propose a simple 
renormalizable left-right symmetric theory where the 
neutrino masses are generated through a double seesaw 
mechanism which is a combination of the Type I and Type III 
seesaw mechanisms. We find a cold dark matter candidate 
which is like the wino in the minimal supersymmetric SM and 
discuss the leptogenesis mechanism. We investigate the 
spectrum of the theory and the possible signals at the LHC.
We refer to this theory as ``Type III-LR".

This work is organized as follows: In section II we discuss the
main features of left-right theories and the different mechanisms
to generate neutrino masses. We show for the first time the
implementation of Type III seesaw. In section III we discuss 
the possible dark matter candidates and the Baryogenesis 
via Leptogenesis mechanisms, while in section IV we 
investigate the spectrum of the theory and the main signals at
future colliders. In section V we summarize our main results.
\section{II. Type III Seesaw in Left-Right Theories}
The so-called left-right symmetric models are one of the 
most appealing extensions of the SM where one can understand 
the origin of parity violation in a simple way and we can generate
neutrino masses. The simplest theories are based on the gauge group
$SU(3)_C \bigotimes SU(2)_L \bigotimes SU(2)_R \bigotimes U(1)_{B-L}$.
Here $B$ and $L$ stand for baryon and lepton number, respectively.
The matter multiples for quarks and leptons are given by
\begin{equation}
Q_L = \left(
\begin{array} {c}
u_L \\ d_L
\end{array}
\right) \ \sim \ (2,1,1/3),
Q_R = \left(
\begin{array} {c}
 u_R \\ d_R
\end{array}
\right) \ \sim \ (1,2,1/3),
\end{equation}
\begin{equation}
l_L = \left(
\begin{array} {c}
 \nu_L \\ e_L
\end{array}
\right) \ \sim \ (2,1,-1),
\end{equation}
and
\begin{equation}
l_R = \left(
\begin{array} {c}
 \nu_R \\ e_R
\end{array}
\right) \ \sim \ (1,2,-1).
\end{equation}
Predicting the existence of right-handed neutrinos.
Here we omit the properties of the multiplets
under $SU(3)_C$. Therefore, one expects
that neutrino masses are generated at least
through the Type I~\cite{TypeI-1,TypeI-2}
seesaw mechanism. Under the left-right parity
transformation one has the following relations
\begin{equation}
Q_L \longleftrightarrow Q_R \ \qquad \text{and} \qquad l_L \longleftrightarrow l_R.
\end{equation}
The relevant Yukawa interactions for quarks in this 
context are given by
\begin{eqnarray}
- {\cal L}_Y^{quarks} &=& \bar{Q}_L  \left( Y_1 \Phi \ + \ Y_2 \tilde{\Phi} \right) Q_R \ + \ \text{h.c.},
\end{eqnarray}
where the bidoublet Higgs is given by
\begin{equation}
\Phi = \left(
\begin{array} {cc}
 \phi_1^0   &  \phi^+_2 \\
 \phi^-_1  & \phi_2^0
\end{array}
\right) \ \sim \ (2, 2, 0), \qquad \ \text{and} \ \qquad \tilde{\Phi}= \sigma_2 \Phi^* \sigma_2.
\end{equation}
Once the bidoublet gets the vev the quark mass matrices read as
\begin{eqnarray}
M_U & = & Y_1 v_1 \ + \ Y_2 v_2^*, \ \text{and} \
M_D =  Y_1 v_2 \ + \ Y_2 v_1^*,
\end{eqnarray}
with $v_1= \langle \phi_1^0 \rangle$, and $v_2 = \langle \phi_2^0 \rangle$.
In the case of the bidoublet one has the following transformation
under the left-right parity
\begin{equation}
\Phi \longleftrightarrow \Phi^\dagger.
\end{equation}
and $Y_1 = Y_1^\dagger$ and $Y_2 = Y_2^\dagger$. These aspects of the model
have been studied in detail~\cite{GoranLR, Senjanovic-Mohapatra}.
Now, as it has been noticed by Mohapatra and Senjanovi\'c, in this context one
can have a deep relation between the origin of neutrino masses and
parity violation~\cite{TypeI-2}. In the most popular renormalizable
left-right models, the neutrino masses are generated through the
Type I~\cite{TypeI-1,TypeI-2} and Type II~\cite{TypeII} seesaw mechanisms
once the Higgs triplets, $\Delta_L \sim (3,1,2)$ and $\Delta_R \sim (1,3,2)$
are introduced. Predicting the existence of doubly charged Higgses which could be
discovered at the LHC~\cite{doublyH}. Before study the implementation of 
Type III seesaw in this context we discuss briefly
the different well known possibilities to generate neutrino masses.
\subsection{II. A. Dirac Neutrinos}
In this context the charged lepton masses
are generated through the interactions
\begin{equation}
-{\cal L}_l= \bar{l}_L \left( Y_3 \ \Phi \ + \ Y_4 \tilde{\Phi} \right) l_R \ + \ \text{h.c.},
\end{equation}
and the relevant mass matrix is given by
\begin{equation}
M_e= Y_3 \ v_2 \ + \ Y_4 v_1^*.
\end{equation}
At the same time one gets a Dirac mass matrix for the neutrinos
\begin{equation}
M_\nu^D= Y_3 \ v_1 \ + \ Y_4 v_2^*.
\end{equation}
In the limit $v_2 \ll v_1$ and $Y_3 \ll Y_4$,
\begin{equation}
M_e \approx  Y_4 \ v_1^*,  \qquad \text{and} \qquad  M_\nu^D= v_1 \left( Y_3 \ + \ M_e \ \frac{v_2^*}{|v_1|^2} \right).
\end{equation}
Therefore, assuming that $Y_3$ is very small one can have Dirac-neutrinos.
However, in this case one has the same situation as in the SM plus
right-handed neutrinos where we can assume a small Dirac Yukawa coupling
for neutrinos. As is well-known in this scenario one has to introduce
extra Higgses in order to break parity and the
left-right symmetry~\cite{GoranLR, Senjanovic-Mohapatra}.
\subsection{II. B. Majorana Neutrinos: Type I plus Type II}
In the so-called minimal left-right theories it is assumed that the neutrino
masses are generated through the Type I and Type II seesaw mechanisms introducing a
pair of Higgs triplets, $\Delta_L \sim (3,1,2)$ and $\Delta_R \sim (1,3,2)$~\cite{TypeI-2}.
In this case the relevant interactions are given by
\begin{equation}
- {\cal L}_\nu= {\cal L}_l \ + \ h \left( l_L^T \ C \ i \sigma_2 \Delta_L \ l_L \ + \ l_R^T \ C \ i \sigma_2 \Delta_R \ l_R \right) \ + \ \text{h.c.},
\end{equation}
$h=h^T$, and
\begin{equation}
\Delta_{L,R} = \left(
\begin{array} {cc}
 \frac{1}{\sqrt{2}}\delta_{L,R}^{+}  &  \delta^{++}_{L,R} \\
 \delta^0_{L,R}  & - \frac{1}{\sqrt{2}} \delta_{L,R}^+
\end{array}
\right).
\end{equation}
Under the left-right parity transformation one has the following relation
\begin{equation}
\Delta_L \longleftrightarrow \Delta_R.
\end{equation}
In this case the mass matrix for neutrinos is given by
\begin{equation}
M_{\nu}^{I-II} = \left(
\begin{array} {cc}
 \sqrt{2} h k_L &  (M_\nu^D)^*  \\
 (M_\nu^D)^\dagger   & - \sqrt{2} h^* k_R^*
\end{array}
\right)
\end{equation}
where $\langle \delta^0_{L,R} \rangle = k_{L,R}/ \sqrt{2}$ and in the
limit when $M_\nu^D \ll h k_R$ one gets
\begin{equation}
M_{\nu_L} =  \sqrt{2} h k_L  \ - \ (M_\nu^D)^\dagger \ M_{\nu_R}^{-1} \ (M_\nu^D)^*,
\end{equation}
and
\begin{equation}
M_{\nu_R} = \sqrt{2} \ h^* k_R^*
\end{equation}
Therefore, one can understand the smallness of the neutrino masses as a consequence 
of large left-right scale $k_R$ (or $M_{W_R}$) since $k_L = \gamma/k_R$~\cite{TypeI-2}.
Even if this possibility is very appealing we do not know which is the mechanism
responsible for neutrino masses and one should explore all possibilities, or at
least the simplest scenarios at tree level. This is the main goal of this article.
\subsection{II. C. The Case of Type III Seesaw}
The realization of the Type III seesaw mechanism has not been
studied in the context of left-right symmetric theories. 
In order to realize this mechanism one has to introduce 
fermionic triplets (one for each family):
\begin{equation}
\rho_L = \frac{1}{2} \left(
\begin{array} {cc}
 \rho_L^0  &  \sqrt{2} \rho^+_L \\
 \sqrt{2} \rho^-_L  & - \rho_L^0
\end{array}
\right) \ \sim \ (3,1,0),
\end{equation}
and
\begin{equation}
\rho_R = \frac{1}{2} \left(
\begin{array} {cc}
 \rho_R^0  &  \sqrt{2} \rho^+_R \\
 \sqrt{2} \rho^-_R  & - \rho_R^0
\end{array}
\right) \ \sim \ (1,3,0),
\end{equation}
and Higgses in the fundamental representation of $SU(2)_L$ and $SU(2)_R$,
respectively.
\begin{equation}
H_L = \left(
\begin{array} {c}
 \phi_L^+ \\ \frac{\phi^0_L \ + \ i \ A_L^0 }{\sqrt{2}}
\end{array}
\right) \ \sim \ (2,1,1),
\end{equation}
and
\begin{equation}
H_R = \left(
\begin{array} {c}
 \phi_R^+ \\ \frac{\phi^0_R \ + \ i \ G_R^0}{\sqrt{2}}
\end{array}
\right) \ \sim \ (1,2,1).
\end{equation}
In this case the relevant interactions are given by
\begin{eqnarray}
- {\cal L}^{III}_\nu &=& {\cal L}_l \ + \ Y_5 \left( l_L^T \ C \ i \sigma_2 \ \rho_L \ H_L
\ + \ \ l_R^T \ C \ i \sigma_2 \ \rho_R \ H_R \right) \nonumber \\
& + & \ M_\rho \ \text{Tr} \left( \rho_L^T \ C \ \rho_L \ + \ \rho_R^T \ C \rho_R \right) \ + \ \text{h.c.}
\end{eqnarray}
Notice that in this case under left-right parity transformation
one has the following relations
\begin{equation}
\rho_L \longleftrightarrow \rho_R \qquad \text{and} \qquad H_L \longleftrightarrow H_R.
\end{equation}
Therefore, once the Higgses $H_L$ and $H_R$ get the vevs, $v_L$ and $v_R$, parity 
is broken spontaneously. In the case when $v_L=0$ and
$v_R \neq 0$, and integrating out the neutral components 
of the fermionic triplets one finds that the mass matrix 
for neutrinos in the basis $\left( (\nu^C)_R, \ \nu_R, \ \rho^0_R \right)$ 
reads as
\begin{equation}
M_{\nu}^{III} = \left(
\begin{array} {ccc}
 0 & M_\nu^D  & 0
\\
(M_\nu^D)^T & 0 & -\frac{Y_5 v_R}{2\sqrt{2}}
\\
0 & -\frac{Y_5^T v_R}{2 \sqrt{2}} & M_{\rho}
\end{array}
\right),
\end{equation}
and the fermionic triplets, $\rho_L$, do not mix having a 
mass matrix equal to $M_\rho$. As one expects the neutrino masses 
are generated through the Type I and Type III seesaw mechanisms
and one has a ``double-seesaw" mechanism since the mass of
the right-handed neutrinos are generated through
the Type III seesaw once we integrate out $\rho_R^0$.

Assuming that $M_\rho \gg Y_5 v_R / 2 \sqrt{2} \gg M_\nu^D$
one gets
\begin{equation}
M_{(\nu^C)_R} = M_\nu^D \ M_{\nu_R}^{-1} \ \left(M_\nu^D \right)^T
\end{equation}
with
\begin{equation}
M_{\nu_R} = \frac{v_R^2}{8} \ Y_5 \ \left( M_{\rho} \right)^{-1} \ Y_5^T.
\end{equation}
Notice the double-seesaw mechanism, where the mass of the right-handed neutrinos
are generated once the fermionic triplet is integrated out (Type III seesaw),
and later the light neutrinos get the mass through the usual Type I mechanism.
Here we stick to the case $v_L=0$ since it has been shown
in~\cite{GoranLR} that this solution corresponds to the
minimum of the scalar potential. Also as we will show in the
next section when $\rho_L \to - \rho_L$, which forbids the mixing between
$\rho_L$ and $l_L$, is a symmetry of the theory the neutral component
of the fermionic triplets can be a cold dark matter candidate.
See Ref.~\cite{Mohapatra-Valle} for an early discussion 
of the double seesaw mechanism in a different context. 

As we have seen in this case since one has a double seesaw mechanism we
can have an interesting scenario for the LHC where the fermionic triplets
are at the TeV scale, $M_{\rho} \approx 1 $ TeV, and the right-handed neutrinos
at the scale, $M_{\nu_R} \approx 10$ GeV. Therefore, as it is well known
in this case one gets small neutrino masses, $m_\nu \approx 1$ eV, if the Yukawa 
couplings are very small. See references~\cite{Strumia, deAguila} for the 
production of fermionic triplets at the LHC. In the case when we assume that 
the left-righ symmetry scale is very large, $v_R \sim 10^{14-15}$ GeV, one
can have a scenario where $Y_5 < 1$, $M_\rho \gg v_R$, $M_{\nu}^D \sim M_W$
and one gets $m_\nu \sim 1$ eV. In the next section we will discuss the first
scenario in order to understand the possibility to test this theory at
future collider experiments.
\section{III. Cold Dark Matter and Leptogenesis}
In this theory a possible candidate for the cold dark matter
in the Universe is the neutral components of $H_L$ once we
impose the symmetry $H_L \to - H_L$. This situation is
similar to the case of Inert Higgs Doublet Models~\cite{IDM}.
At tree level the charged component, $\phi_L^+$, and
the neutral components, $\phi_L^0$ and $A_L^0$ in $H_L$
have the same mass. However, as it has been pointed
out in Ref.~\cite{strumia} once the radiative corrections
are considered the charged component is heavier and
the neutral components could be a good cold dark matter
candidate. Unfortunately, in this case since the real and imaginary
components have the same mass one cannot satisfy the
constraints coming from direct detection. Then, we do not
stick to this possibility. It is important to emphasize that the neutral
component of $H_R$ cannot be a CDM candidate since one has to
break parity and the left-right symmetry, and in the case of
the neutral components of the bidoublet one knows
that both vevs should be different from zero since one
has to generate fermion masses in agreement with
the experiment.

In the fermionic sector one can find a natural
cold dark matter once we impose the symmetry
$\rho_L \to - \rho_L$. In this case the cold dark
matter candidate is the neutral component of the
lightest fermionic triplet $\rho_L$.
Since in this theory one has to introduce three fermionic
triplets, one for each generation, there are three fields
which have the same quantum numbers as the winos in
the minimal supersymmetric SM. At tree level the charged
and neutral components of the lightest triplet have
the same mass. However, once the radiative corrections
are included one has the splitting $\Delta M \approx 166$
MeV and the charged component decays into the dark matter
and a pion before nucleo-syntesis. A cold dark matter
candidate with the same quantum numbers has been studied
in~\cite{strumia} where the authors pointed out
that in order to explain the CDM relic density
the mass has to be $M_{\rho_L} \approx 2.5$ TeV.
Since in our case one has three fermionic triplets
the cold dark matter candidate has to be the
lightest neutral component, $\rho_L^0$.

In this theory one could have in principle different
leptogenesis mechanisms. In the case of the natural
double seesaw mechanism discussed in the previous section
where one has $M_\rho \gg Y_5 v_R / 2 \sqrt{2} \gg M_\nu^D$,
the right-handed neutrinos are lighter than the fermionic
triplets. Therefore, one has the so-called Type I
Leptogenesis~\cite{Fukugita} with extra vertex corrections where 
we have the fermionic triplets inside the loops. In Ref.~\cite{Steve} 
it has been studied the leptogenesis mechanism when the neutrino
masses are generated through the Type I and Type III seesaw mechanisms 
and it has been shown that the fermionic triplets cannot contribute to the
self-energy corrections. These issues will be studied in detail in
a future publication.
\section{IV. Spectrum of the Theory and Possible Signals at the LHC}
In the previous section we have discussed the main
properties of the theory. The properties of a left-right theory 
where the Higgs sector is composed of $\Phi$,
$H_L$ and $H_R$ have been studied in detail
by Senjanovi\'c~\cite{GoranLR} in a seminal paper.
In this context the mass matrix for the gauge bosons
$W_L^{\pm}$, and $W_R^{\pm}$ when $v_L=0$ is
given by~\cite{GoranLR}:
\begin{equation}
{\cal M}_{\pm}^2 =
\left( \begin{array} {cc}
 \frac{g^2}{4} (v_1^2 + v_2^2) & - \frac{g^2}{2} v_1 v_2
 \\
- \frac{g^2}{2} v_1 v_2 &  \frac{g^2}{4} (v_1^2 + v_2^2 + v_R^2)
\end{array} \right),
\end{equation}
where $g_L=g_R=g$ and the mass matrix for the neutral gauge bosons,
$Z$, $Z^{'}$ and $A$ reads as
\begin{equation}
{\cal M}_{0}^2 =
\left( \begin{array} {ccc}
 \frac{g^2}{4} (v_1^2 + v_2^2) & - \frac{g^2}{2} (v_1^2 + v_2^2) & 0
 \\
- \frac{g^2}{2} (v_1^2 + v_2^2) &  \frac{g^2}{4} (v_1^2 + v_2^2 + v_R^2) & -\frac{1}{4} g \tilde{g} v_R^2
\\
0 & -\frac{1}{4} g \tilde{g} v_R^2 & \frac{1}{4} \tilde{g}^2 v_R^2
\end{array} \right),
\end{equation}
with $\tilde{g}$ being the gauge coupling
for the $U(1)_{B-L}$ symmetry.
In the case of the Higgs bosons when $v_L=0$ only
the Higgses in $\Phi$ and $H_R$ mix once the symmetry is broken.
Notice that in these two fields one has 12 degrees of freedom,
six Goldstone bosons and six physical Higgs bosons in the theory~\cite{GoranLR}.
Therefore, in total one has ten physical Higgses, four charged Higgses
$H_1^{\pm}$ and $H_2^{\pm}$, four CP-even neutral Higgses $H_1^0, H_2^0, H_3^0$
and $H_4^0$, and two CP-odd states $A_1^0$ and $A_2^0$.

Let us discuss the possible signals at the LHC which could help
us to identify this theory. As usual in this context one predicts the
existence of extra gauge bosons, $W_R^{\pm}$ and $Z^{'}$. As is well-known
the discovery of these states is crucial to test any left-right symmetric theory.
In the case of $W_R$ one has the mechanism at the LHC,
$pp \to W_R^* \to e_R \nu_R$, where $\nu_R$ decays into $e_R j j$~\cite{Krasnikov}.
Then, one has two leptons and two jets with high $p_T$. Defining the invariant
mass $M_{inv} (ejj)$ and $M_{inv}(eejj)$ for $\nu_R$ and $W_R$, respectively,
one should be able to make the reconstruction and identify these fields.
In the case of the $Z^{'}$~\cite{Frank} one has the production
mechanism, $pp \to (Z^{'})^* \to e^+ e^-$ and one looks for excess
of dilepton events.

Once we implement the Type III seesaw mechanism in this context one finds
that parity conservation at the left-right scale tells us that the masses of
the fermionic triplets $\rho_L$ and $\rho_R$ should be the same. This could be
a way to test the model at the LHC if one finds these states and determines
their masses. At the same time since the neutrino masses are generated through the
Type I and Type III seesaw mechanisms one needs to discover the right-handed
neutrinos~\cite{Tao, deAguila} and the fermionic triplets to test the theory. Now,
in the case of the fermionic triplets $\rho_L$ one has the
following production mechanisms, $pp \to Z^*, \gamma^*, (Z^{'})^* \to \rho_L^+ \rho^-_L$
and $pp \to W_L^* \to \rho_L^{\pm} \rho_L^0$. Now, if we stick to
the possibility that the neutral component is responsible
for the CDM,  the charged component will have a decay length
of few centimeters and decay into the neutral component and a pion.
The fermionic triplets $\rho_R$ can be produced via
$pp \to \gamma^*, (Z^{'})^* \to \rho_R^+ \rho^-_R$ or
$pp \to W_R^* \to \rho_R^{\pm} \rho_R^0$. In this
case $\rho_R^0$ could decay into a lepton and two jets,
while $\rho^{+}_R$ decays into three leptons.
All these issues, and the constraints on the $W_R$ mass and the Higgs masses
coming from low-energy processes will be studied in a future publication.
\section{V. Summary and Outlook}
The implementation of the Type III seesaw mechanism for
neutrino masses in the context of left-right symmetric 
theories where parity is broken spontaneously has been 
investigated. We have presented a simple left-right symmetric 
theory where the neutrinos masses are generated through a ``double
seesaw" mechanism which is a combination of the Type I and Type III 
seesaw mechanisms. We have found that the lightest neutral component 
of the fermionic triplets $\rho_L$ can be a candidate for the 
cold dark matter in the Universe. In this context one could have 
Type I leptogenesis with extra vertex contributions due to the existence
of the fermionic triplets inside the loops. We have discussed the 
spectrum of the theory, and the possibilities to realize the test 
at the LHC. The phenomenological and cosmological aspects of this
proposal are very rich and deserve to be investigated.
\\
\\
\textit{Acknowledgments}. I would like to thank I. Dorsner, M. Drees 
and G. Senjanovi\'c for the careful reading of the manuscript and discussions, 
and T. Han for discussions. This work was supported in part by the U.S.
Department of Energy contract No. DE-FG02-08ER41531 and in part by the
Wisconsin Alumni Research Foundation.

\end{document}